\documentclass[sigconf]{acmart}

\usepackage{tabularx} 
\usepackage[flushleft]{threeparttable} 
\usepackage{multirow}
\usepackage{enumitem}
\usepackage{balance}

\AtBeginDocument{%
  \providecommand\BibTeX{{%
    \normalfont B\kern-0.5em{\scshape i\kern-0.25em b}\kern-0.8em\TeX}}}

\copyrightyear{2020} 
\acmYear{2020} 
\setcopyright{acmlicensed}\acmConference[CHI PLAY '20 EA]{Extended Abstracts of the 2020 Annual Symposium on Computer-Human Interaction in Play}{November 2--4, 2020}{Virtual Event, Canada}
\acmBooktitle{Extended Abstracts of the 2020 Annual Symposium on Computer-Human Interaction in Play (CHI PLAY '20 EA), November 2--4, 2020, Virtual Event, Canada}
\acmPrice{15.00}
\acmDOI{10.1145/3383668.3419911}
\acmISBN{978-1-4503-7587-0/20/11}

\begin{document}
\fancyhead{} 

\title{Playing With Friends --- The Importance of Social Play During the COVID-19 Pandemic}

\author{Sebastian Cmentowski}
\email{sebastian.cmentowski@uni-due.de}
\affiliation{%
  \department{High-Performance Computing}
  \institution{University of Duisburg-Essen}
  \city{Duisburg}
  \country{Germany}
}

\author{Jens Kr\"uger}
\email{jens.krueger@uni-due.de}
\affiliation{%
\department{High-Performance Computing}
  \institution{University of Duisburg-Essen}
  \city{Duisburg}
  \country{Germany}
}

\renewcommand{\shortauthors}{Cmentowski and Kr\"uger}

\begin{abstract}
In early 2020, the virus SARS-CoV-2 evolved into a new pandemic, forcing governments worldwide to establish social distancing measures. Consequently, people had to switch to online media, such as social networks or videotelephony, to keep in touch with friends and family. In this context, online games, combining entertainment with social interactions, also experienced a notable growth. In our work, we focused on the potential of games as a replacement for social contacts in the COVID-19 crisis. Our online survey results indicate that the value of games for social needs depends on individual gaming habits. Participants playing mostly multiplayer games increased their playtime and mentioned social play as a key motivator. Contrarily, non-players were not motivated to add games as communication channels. We deduce that such crises mainly catalyze existing gaming habits.

\end{abstract}

\ccsdesc[500]{Applied computing~Computer games}
\ccsdesc[500]{Human-centered computing~Empirical studies in collaborative and social computing}

\keywords{games; social play; COVID-19; social interactions}

\maketitle

\section{Introduction}
In the first months of 2020, the world was struck by a new pandemic. The virus SARS-CoV-2, causing the respiratory disease COVID-19, first appeared in the Chinese province Hubei and quickly spread worldwide. As a consequence of exponentially rising infection rates, most countries imposed severe measures. The goal was to slow the spread of the virus and protect the healthcare systems. These actions were mostly aimed at preventing situations in which people could be close enough for droplet infection. The individual measures to achieve social distancing ranged from using smaller working shifts or home office to drastic steps such as curfews.

Even though the infection curve in many countries was successfully flattened, the lockdown of public life had widespread consequences. Above all, staying at home and limiting personal contacts to a minimum reduced social interactions to online platforms. It is no surprise that software providers for video-conferences, chats, or social networks reported significant consumer growth. However, this trend was also observed for another communication-intensive media type: online gaming. For example, Verizon reported a 75\% increase in gaming usage in peak hours for the United States~\cite{shanley_GamingUsage}.

This trend could indicate that more people are using digital games as communication media to keep in touch with their friends. For instance, Jones describes her gaming activities in her auto-ethnography as an opportunity to "cement social engagement with a wider group of friends"~\cite{jones_2020}. Even though social play~\cite{isbister2010enabling} is a well-researched topic, the unique circumstances of the worldwide isolation measures provide an ideal situation to examine the effects on gaming habits.

\begin{table*} 
    \caption{Custom Questions (CQ) used in the online survey. Answers are either provided in free-text form or on a fixed scale.} 
  \begin{tabularx}{\textwidth}{
  p{1cm}  X  p{2cm}}
    \toprule 
   \multicolumn{2}{l}{\textbf{Question}}& \textbf{Type} \\
    \midrule
    \textbf{RQ1:}&\textbf{Does the COVID-19 pandemic influence gaming habits?} \\
     CQ1: & How would you describe your gaming habits before the COVID-19 pandemic? & 5-point scale\\
     CQ2: & Did you change your gaming habits in the last three months? & 6-point scale\\
     CQ3: & If your gaming habits changed, please state your personal reasons. & Free-text form\\
     \addlinespace
     \textbf{RQ2:}&\textbf{Is there a shift regarding the popular genres?} \\
     CQ4: & Which games did you play more than 10hrs in 2019? & Free-text form\\
     CQ5: & Which games did you play within the last 3 months? & Free-text form\\
     CQ6: & What were your main reasons for choosing these specific games during the last 3 months? & Free-text form\\
     \addlinespace
     \textbf{RQ3}&\multicolumn{2}{l}{\textbf{How important do players rate social interactions in games during the pandemic?}} \\
     CQ7: & What are your personal main reasons to play games?  & Free-text form\\
     CQ8: & How important would you rate social interactions ( collaborative or competitive tasks) in games? & 5-point scale\\
     CQ9: & Do you think that your rating on social interactions is biased by the current lockdown situation?  & 5-point scale\\
  \bottomrule
\end{tabularx}
\label{tab:customQuestions}
\end{table*}

To this end, our research addresses the question: How important are social activities in games for players affected by isolation measures? We subdivide this question into three parts, forming a holistic image of how the crisis has changed the personal gaming behavior, focusing on social interactions. First, we assess the influence of the pandemic on gaming habits. Next, we examine whether the circumstances provoked a shift in popular games or genres. Finally, we ask players to rate social interactions in games directly.

We conducted an online study analyzing the players' gaming preferences and habits within the first months of 2020. The results indicate that the COVID-19 crisis influenced the gaming behavior of active gamers. Roughly 50\% of players increased their gaming activities, played more multiplayer titles, and listed social aspects as their key motivators. However, these effects are mostly limited to the gamer subgroup using multiplayer games as a social medium. On the other hand, our study implies that the crisis did not motivate new players to include games as communication channels.

\section{Related Work}
Gaming activities can fulfill a wide variety of personal needs, e.g., enjoyment, social interactions, or challenging situations~\cite{tondelloPlayer}. Especially in global crises, games can help cope with the negative effects of social isolation~\cite{oe2020discussion}. For instance, exergames could help replace exercise facilities or outdoor activities and aid in treating anxiety disorders~\cite{viana2020exergames}. Besides, multiplayer games are known to strengthen the connectedness with a virtual community~\cite{deppingMultiplayer}, which could benefit the overall mental health by reducing feelings of solitude. However, increased gaming activities may also intensify the downsides in the form of addictive effects. King et al.~\cite{king2020problematic} discuss the risks of online gaming in the context of COVID-19 and mention a potential increase in gaming disorders paired with a decrease in social connections. Nevertheless, at the same time, the closure of casinos and other gaming facilities might mitigate the overall situation ~\cite{ruddock2020life}.

In this work, we place a particular focus on the social play aspect that can help people cope with reduced social activity. Social play~\cite{isbister2010enabling} covers a wide variety of settings in which multiple people are involved in a single game. These include both co-located sessions and distributed games. In the past, social play has been thoroughly covered in research~\cite{paavilainen2013social, segura2015enabling}. In particular, work focused on the inter-player relations and interactions that shape the social context between the players~\cite{emmerich2013helping, harris2019asymmetry}. A key concept to the player experience is the social presence~\cite{biocca2003toward, de2008people, jansz2005gaming}, described as "the sense of being together with another"~\cite{biocca2003toward}. Games eliciting such feelings can reduce the sense of solitude caused by social distancing.

\section{Online Survey}
Our research's central goal was to determine how the COVID-19 pandemic influenced the players' gaming preferences and habits. Therefore, we conducted a qualitative online study assessing the participants' personal situations and opinions. The final questionnaire comprised three custom questions per RQ and three questions assessing the demographics of the participants (see Table~\ref{tab:customQuestions}). Since we were interested in how the general public played games during the lockdown, we refrained from targeting specific gaming communities, e.g., Discord chats dedicated to a single title. Instead, we promoted the survey on general online channels, such as Twitter or Reddit. Additionally, we used some online groups dedicated to the recruitment of survey participants.

In total, 78 participants completed the survey. Most subjects were of German nationality (65 participants), followed by American (7), Austrian (2), and Belgian (4) nationality. The mean age was 25.4 (\textit{SD}~=~6.79), with a range from 16 to 57. Concerning the gender, the sample was slightly uneven: 35 males and 43 females participated in our study. 

\begin{figure*}[t!]
\centering
\includegraphics[width=0.93\textwidth]{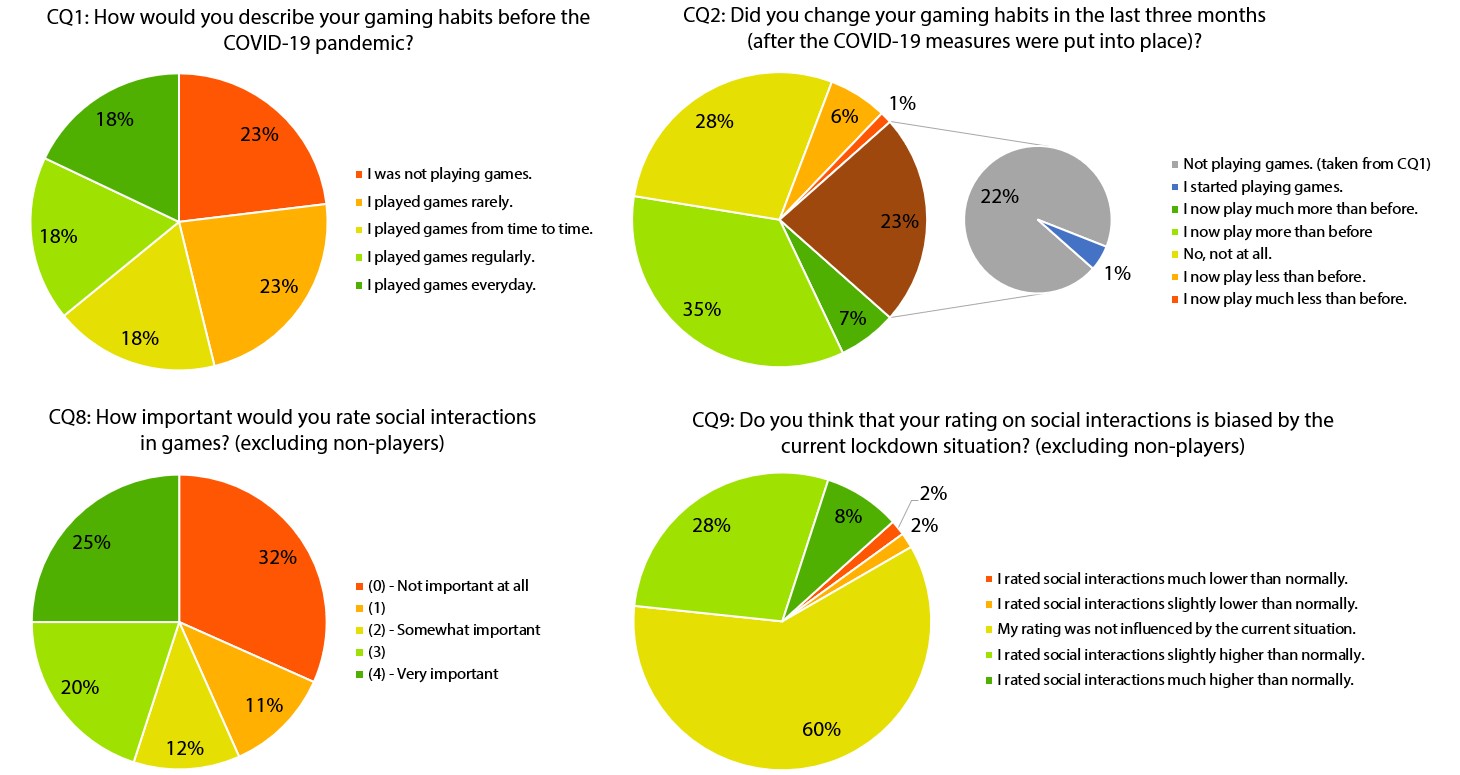}
\caption{We asked the subjects several questions about their gaming habits and their opinion on social interactions. The pie charts illustrate the respective questions CQ1, CQ2, CQ8, CQ9, and the overall answer distribution of the 78 participants.
}
\label{fig:gaminghabits}
\end{figure*}

\begin{table} 
    \caption{The participants listed various reasons for playing more/less during the COVID-19 crisis.} 
  \begin{tabularx}{\columnwidth}{
    p{0.1cm} X >{\raggedleft\arraybackslash} p{2.3cm}}
    \toprule 
   \multicolumn{2}{l}{\textbf{Reasons for playing more}}& \multicolumn{1}{r}{\textbf{Frequency}} \\
    \midrule
     \multicolumn{2}{l}{\textbf{more time}, e.g., due to home office}& \textbf{46.27\%} \\
    \multicolumn{2}{l}{\textbf{social needs}, e.g., meet friends virtually}& \textbf{35.82\%} \\
    \multicolumn{2}{l}{\textbf{lack of alternative activities / boredom}}& \textbf{11.94\%} \\
    \multicolumn{2}{l}{\textbf{other reasons}, e.g., new game releases}& \textbf{5.97\%} \\
    \toprule
    \multicolumn{2}{l}{\textbf{Reasons for playing less}}& \multicolumn{1}{r}{\textbf{Frequency}} \\
    \midrule
    \multicolumn{2}{l}{\textbf{less time}, e.g., occupied with other duties}& \textbf{71.43\%} \\
    \multicolumn{2}{l}{\textbf{not in the mood}}& \textbf{28.57\%} \\
  \bottomrule
\end{tabularx}
\label{tab:reasonsHabits}
\end{table}

\section{Results}

The first questions dealt with personal gaming habits. When asked for their gaming behavior before the pandemic (CQ1), the subjects split almost equally between the five options, ranging from everyday gamers to non-players (cf. Figure~\ref{fig:gaminghabits}). The participants were asked whether their habits changed during the crisis. Only one subject started playing during this period. Excluding the 23\% non-players, a majority of 53.5\% played more than before the pandemic, 36.7\% did not alter their gaming behavior, and 10\% reduced their playtime. Asking for the reasons behind these decisions (cf. Table~\ref{tab:reasonsHabits}), most participants reported having more time or an increased need for social contacts. Subjects reducing the playtime were occupied with other duties or not in the mood for games.

In the second part, we asked for the played games before (CQ4) and during the crisis (CQ5), which resulted in two lists of 96 and 91 games. The most popular titles were similar in both cases, e.g., \textit{Counter Strike: Global Offensive}~\cite{CSGO}, \textit{League of Legends}~\cite{LoL}, or \textit{Rocket League}~\cite{RL}. The only differences were due to novel releases, such as \textit{Valorant}~\cite{Valorant} or \textit{Animal Crossing: New Horizons}~\cite{AnimalCrossing}. Also, we could not find any significant difference in popular genres, which were led by First-Person Shooters, Action-Adventures, Sports, and Social Simulations, but did not differ by more than four percent. However, we measured a general shift towards games with multiplayer capabilities. In CQ4, 68.4\% of the listed games offered multiplayer modes. This portion increased by 9.9\% to 78.3\% for the second question (CQ5). Meanwhile, the percentage of single-player games decreased from 70.8\% to 63.1\%.

In CQ6, we asked the participants to list reasons for their game choices during the pandemic. CQ7 inquired about the reasons for playing games in general. In both questions, we collected 29 distinct reasons, including common answers and particularities, e.g., using games as physical workouts. We grouped the reasons into seven categories: fun/entertainment, game characteristics, playing with friends, distraction, challenges, commitment, and external factors. 

\begin{figure}[t]
\centering
\includegraphics[width=1.0\columnwidth]{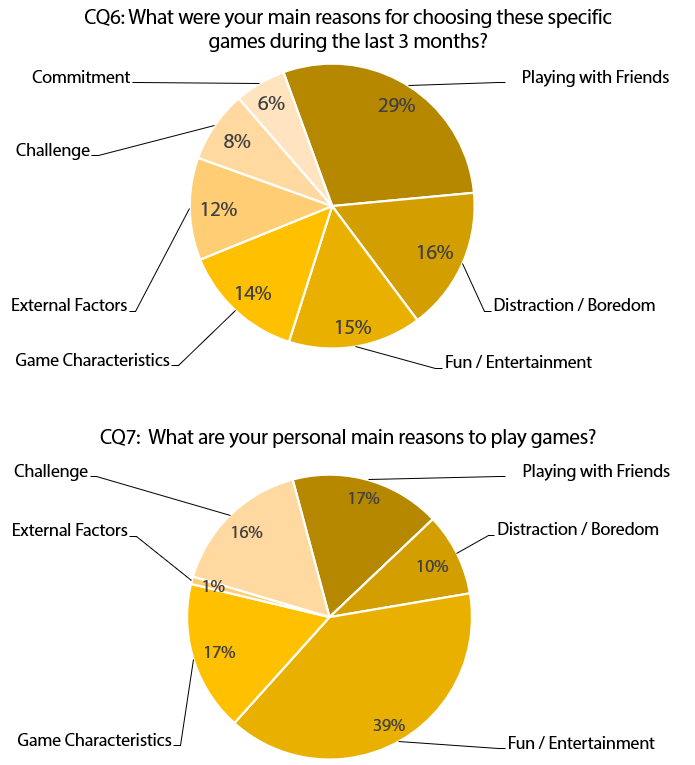}
\caption{We asked the participants to list their reasons for playing games in general (bottom, CQ7) and for choosing specific games during the COVID-19 crisis (top, CQ6).}
\label{fig:gamereasons}
\end{figure}

Between CQ6 and CQ7, we identified major differences depicted in Figure~\ref{fig:gamereasons}. Before the COVID-19 pandemic, the subjects reported playing primarily for fun, game characteristics, and social contacts. Even though these reasons remained relevant in the last months, the emphasis changed. Now, the ranking is dominated by the social aspect of playing with friends, and the distraction from everyday life.

In the final two questions, we asked the participants to rate the importance of social interactions directly and asked them whether the ongoing lockdown measures biased this rating (cf. Figure~\ref{fig:gaminghabits}). Overall, the participants had a very diverse opinion on social activities. 45\% rated these interactions as rather important, 43\% as mostly unimportant. Only a third of the subjects stated that the current situation positively altered their rating, while the majority claimed no influence.

\section{Discussion}

\subsubsection*{RQ1: Does the pandemic influence gaming habits?}\hfill\\
The vast majority of participants (77\%) reported playing games at least once or twice per month, while 23\% never played games. Only one participant started playing within the crisis, to replace physical outdoor activities. Meanwhile, 53.5\% of the gamers increased their playtime in the last months, mostly due to having more free time and an increased need for social contacts. These observations illustrate that while the current circumstances have not encouraged non-players to add gaming to their leisure activities, they have led to an increased playing behavior among gamers.

\subsubsection*{RQ2: Is there a shift regarding popular game genres?}\hfill\\
We were not able to identify big changes in the played game genres. However, the games played before and during the pandemic differed in the ratio of single- to multiplayer games. A possible explanation for the nearly 10\% increase in multiplayer titles could be derived from the assessed reasons to choose these games. The biggest motivators reported by the participants were the social connection when playing with friends, followed by the distraction from frightening circumstances in their everyday life.
 
\subsubsection*{RQ3: How important do players rate social interactions in games during the pandemic?}\hfill\\
When being asked for the importance of social interactions in games, only 25\% of the subjects rated them very important, while 34\% stated valuing these social connections more important than before the crisis. This observation does not contradict our other findings, where social play with friends is the most important motivator to increase the own gaming behavior and play specific titles. Despite a large number of multiplayer games, only 36\% were sole online coop or competitive titles. These were mainly listed by participants rating social interactions as important. Additionally, these subjects also named social play as their central motivator and mostly reported increasing their playtime. Thus, it seems that only a limited subset of players use games actively as a means of socialization. This subset is more likely to increase gaming activities to replace the social contacts reduced by protective measures. Other players, e.g., those preferring single-player games, do not profit in the same way and do not consider games important for social communication.

\section{Conclusion}

Global crises, like the COVID-19 pandemic, fuel the demand for online communication channels, preserving social contacts and replacing face-to-face encounters. Apart from social networks and videotelephony, digital games are another potential source for fulfilling individual social needs. Our research addressed the question: How important are social activities in games for players affected by isolation measures? 

Our online survey results indicate that the value of games for social needs depends on the individual gaming habits. The subset of participants playing mostly multiplayer games increased their playtime and mentioned social play as a key motivator. In turn, non-players were not motivated to use games as another communication channel. Therefore, it seems that crises like COVID-19 mainly catalyze existing habits.

Our results are based upon a limited set of subjects, mostly residing in one country, who were asked after the first wave of infection. While this phase was very similar in most countries, the pandemic's subsequent course varied greatly across the globe. The varying infection rates and countermeasures provide an ideal setting for further studies assessing the long-term impact of isolation measures on online media use.

\bibliographystyle{ACM-Reference-Format}
\balance
\bibliography{literature}

\end{document}